\documentstyle[12pt,aasms4,flushrt]{article}

\newcommand{\aliii}{Al~{\sc iii}}
\newcommand{\ciii}{C~{\sc iii}]}
\newcommand{\civ}{C~{\sc iv}}
\newcommand{\feii}{Fe~{\sc ii}}
\newcommand{\feiii}{Fe~{\sc iii}}
\newcommand{\mgii}{Mg~{\sc ii}}
\newcommand{\oii}{[O~{\sc ii}]}
\newcommand{\oiii}{[O~{\sc iii}]}
\newcommand{\siiv}{Si~{\sc iv}}
\def\ha{\ifmmode {{\rm H}\alpha}
	\else {H$\alpha$}\fi}
\def\hb{\ifmmode {{\rm H}\beta}
	\else {H$\beta$}\fi}

\begin{document}
\title{HAWAII 167 and Q0059-2735: \\
Heavily Dust-Enshrouded Young QSOs}

\author{E. Egami\altaffilmark{1,2}}
\affil{Institute for Astronomy, University of Hawaii, 2680 Woodlawn
Drive, Honolulu, HI 96822\\
egami@mpe-garching.mpg.de}

\author{F. Iwamuro, T. Maihara, and S. Oya}
\affil{Department of Physics, Kyoto University, Kitashirakawa, Kyoto
606-01, Japan\\
iwamuro@cr.scphys.kyoto-u.ac.jp, maihara@cr.scphys.kyoto-u.ac.jp,
oya@cr.scphys.kyoto-u.ac.jp } 

\and

\author{L. L. Cowie}
\affil{Institute for Astronomy, University of Hawaii, 2680 Woodlawn
Drive, Honolulu, HI 96822\\
cowie@ifa.hawaii.edu}

\altaffiltext{1}{Visiting Astronomer,
United Kingdom Infrared Telescope, operated by the Royal Observatory
Edinburgh for the UK Science and Engineering Research Council.}
\altaffiltext{2}{Current address: Max-Planck-Institut f\"{u}r
extraterrestrische Physik, Postfach 1603, 85740 Garching, Germany}
\begin{abstract}
Using the OH-airglow suppressor spectrograph at the University of
Hawaii 2.2m telescope and the CGS4 spectrometer at the United Kingdom
Infrared Telescope, we have found exceptionally large Balmer
decrements in two unusual high-$z$ QSOs, Hawaii 167 ($z=2.36$, $\ha /
\hb = 13$) and Q0059-2735 ($z=1.59$, $\ha / \hb = 7.6$), the latter
being a so-called low-ionization broad absorption line QSO (BALQSO).
We argue that these objects are young QSOs heavily enshrouded by dust.
In fact, the internal reddening might be so large as to completely
extinguish the QSO light in the restframe UV, allowing us to see the
underlying stellar population.  Our possible detection of the 4000
\AA\ break in Hawaii 167 supports this idea.  Its small amplitude
indicates a very young age for the population, $\sim$ 15 Myrs.  To
explain the properties of these QSOs, we propose a model in which a
young QSO is surrounded by a shell of young massive stars mixed with
significant amounts of dust.  We predict that as the QSO emerges from
this dust cocoon, it will eventually take on the appearance of a normal
BALQSO.
\end{abstract}

\section{INTRODUCTION}
In the Hawaii deep $K$-band survey (\cite{Cowie94a};
\cite{Songaila94}), one extraordinary object was found at a redshift%
\footnote{The redshift was originally reported as 2.35, but our
reanalysis of the \ha\ line indicates that it is closer to 2.36.}of 2.36, the
highest by far in the sample.  This object, named Hawaii 167, has
strange characteristics: its high redshift, compact morphology, and
broad ($\sim$ 5000 km s$^{-1}$) Balmer emission lines are sure signs
of this object's being a QSO while its restframe UV spectrum resembles
those of starburst galaxies, showing strong metal absorption lines
with no obvious emission features (\cite{Cowie94b}).  Cowie et al.\
(1994b) suggested that Hawaii 167 was either a very exotic broad
absorption line QSO (BALQSO), a starburst galaxy, or most probably a
mixture of both.  If Hawaii 167 is in any way related to starbursts,
it may be of considerable importance: the small area coverage of the
survey ($\sim$ 77 arcmin$^{-2}$) suggests that this type of object
might in fact be quite common in faint near-IR samples.

The key to understanding Hawaii 167 may lie in the QSO called
Q0059-2735 (\cite{Hazard87}), the only previously known object with
similar characteristics.  This object is the most extreme member of
the low-ionization BALQSOs (also often called \mgii\ BALQSOs)
comprising $\sim$ 15\% of BALQSOs, which themselves represent $\sim$ 10\%
of all QSOs (\cite{Weymann91}).  This class of QSOs shows broad
absorption lines (BALs) of low-ionization ions (\mgii , \aliii ) as
well as those of the high-ionization species (\civ , \siiv ) regularly
seen in BALQSOs.  What separates low-ionization BALQSOs from the
normal BALQSOs seems to be their richness in dust: they are known to
be more common in IRAS-selected samples (\cite{Low89}), and their UV
continuum shows signs of moderate reddening (\cite{Sprayberry92}).
Also, the absence or the extreme weakness of \oiii\ in these objects
can be understood if dust is preventing ionizing radiation from
reaching the outer low-density regions where the formation of the
forbidden line is possible (\cite{Boroson92}).  Voit et al.\ (1993)
suggested that low-ionization BALQSOs are probably ``young quasars in
the act of casting off their cocoons of gas and dust,'' which implies
some kind of connection between these objects and the ultraluminous
IRAS galaxies (\cite{Sanders88}).  Indeed, one of the ultraluminous
IRAS galaxies, IRAS 07598+6508, was found to be a low-ionization
BALQSO (\cite{Lipari94}).

As already mentioned, one interpretation for Hawaii 167 put forth by
Cowie et al. (1994b) was that Hawaii 167 is a mixture of both a QSO
and a starburst galaxy: its internal reddening is so large that its
QSO light, dominating the restframe optical, is completely
extinguished in the restframe UV, leaving only the light of the
surrounding starbursting galaxy.  This idea is supported by the lack
of any UV broad emission lines (BELs), and the large Balmer decrement
($> 8$) inferred from the non-detection of \hb.  In this paper, we
will investigate this possibility by examining the near-IR spectra of
Hawaii 167 and Q0059-2735, using the Balmer decrement as an indicator
of internal reddening.
\section{THE DATA}
The $K$-band (\ha ) spectrum of Hawaii 167 was taken on the night of
1993 December 21 using the CGS4 spectrometer (\cite{Mountain90}) at
the United Kingdom Infrared Telescope (UKIRT) on Mauna Kea.  The 75
line mm$^{-1}$ grating used in first order gave a resolving power, R,
of $\sim 340$ with a 3\arcsec\ (1 pixel) slit.  An adequate sampling
was achieved by shifting the detector over 2 pixels in 4 steps in the
wavelength direction.  The integration time at each detector position
was 30 seconds, and the total integration time was 28 minutes.  This
data was summarized in Cowie et al. (1994b).

The $J$-band (\oii ) and $H$-band (\hb ) spectra of Hawaii 167
were taken on the nights of 1994 October 24--30 using the OH-airglow
suppressor spectrograph (\cite{Iwamuro94}) at the University of Hawaii
2.2 m telescope on Mauna Kea.  This spectrograph produces both $J$-
and $H$-band spectra simultaneously with R $\sim$ 100.  With a
1.5\arcsec\ wide slit, this spectrograph first projects $J$- and $H$-band
spectra on a mask mirror with R $\sim$ 5500.  The mask mirror has
narrow black (non-reflecting) stripes on the positions of the OH sky
lines, and therefore removes the contribution from these lines upon
the reflection of the light.  This light is subsequently recombined to
produce final low-resolution spectra (R $\sim$ 100) with a sky
background reduced by a factor of $\sim$ 20.  The total integration
time was 11 hours and 44 minutes (8 minutes $\times$ 88 frames).

Spectra of Q0059-2735 were taken in the $H$ band (\ha ) and $J$ band
(\hb ) on the nights of 1994 October 8 and 9, respectively, using the
CGS4 spectrometer at UKIRT.  This object was observed as a filler
during the observing run allocated for the follow-up spectroscopy of a
$z>4$ galaxy survey (Egami 1995).  The 75 line
mm$^{-1}$ grating was used in first order in the $H$ band and in second
order in the $J$ band. With a 3\arcsec\ (2 pixels) slit, this gave R $\sim
250$ ($H$) and $\sim 380$ ($J$).  The detector was shifted over 2
pixels in 6 steps.  The integration time at each position was 60
seconds, and the total integration time was 36 minutes in each band.

The reduced spectra are shown in Fig.~\ref{fig1} (Hawaii 167) and
Fig.~\ref{fig2} (Q0059-2735).  The measured line parameters are
summarized in Table~1.  These measurements are based on gaussian
profile fitting, and the derived fits are drawn in the figures as the
thick solid lines.  We have adopted the fitting procedure of
Baker et al.\ (1994) so that we can compare our Balmer decrement
measurements with theirs.  The quoted errors are internal fitting
errors (1 $\sigma$) and determined by performing one hundred Monte
Carlo simulations.  By comparing with the broad-band magnitudes listed
in Cowie et al.\ (1994b) (the horizontal dash-dot lines in the
figures), we estimate the photometric accuracy to be 10--20\% after
the contribution from the emission lines is taken into account.  The
broad-band $J$ magnitude of Q0059-2735 seems to be significantly
higher than the continuum level of our spectrum, but we think this
offset could be real.  There are complexes of \feii\ emission lines
just outside our spectral band in both directions, and these \feii\
complexes have been found to be extremely strong in ultraluminous IRAS
galaxies (\cite{Lipari93}) and high-$z$ QSOs (\cite{Hill93}).  These
\feii\ emission complexes are within the passband of the $J$ band, and
might be boosting the broad-band flux level.  In the case of
Q0059-2735, two gaussians were used simultaneously for the \ha\ to
obtain a good fit, and for the \hb\ we mirror-reflected the left half
of the line and then fitted a gaussian in order to avoid the
contamination from \feii\ emission, whose presence is clearly seen in
the figure on the right side of the line.  In the case of Hawaii 167,
we do not think that the peak near 5007 \AA\ is a real feature because
its width was found to be smaller than the instrumental profile; it
also lies very close to the edge of the spectrum, reinforcing the
possibility that this feature is an artifact.
\section{DISCUSSION}
The values of the Balmer decrements found for these QSOs are
exceptionally large.  Although the Balmer decrements of most QSOs are
known to be significantly larger than 2.87, which is the case B
recombination value at T = 10$^{4}$ K, their typical range seems to be
3 -- 6 (\cite{Baker94}).  This is still much smaller than the values
for Hawaii 167 (13) and Q0059-2735 (7.6).  Considering the mounting
evidence for the low-ionization BALQSOs being dust rich, it seems
natural to think that the presently measured large Balmer decrements
are a result of heavy reddening by internal dust extinction.

The amounts of internal reddening inferred from the Balmer decrements
are very significant.  For example, if we assume $\ha / \hb \leq 6$
for normal QSOs, we obtain $E(B-V) \geq 0.19$ and 0.54 for Q0059-2735
and Hawaii 167, respectively, after the foreground Galactic reddening
($E(B-V)=0.02$ for Q0059-2735 and 0.14 for Hawaii 167) is taken into
account.  The actual values may be closer to $E(B-V) = 0.35$ and 0.70,
corresponding to an intrinsic Balmer decrement of 5.  This means that
these QSOs must be heavily enshrouded by dust --- for an extinction
law similar to that in our galaxy, these values of the color excess
correspond to A(1500 \AA ) $\sim$ 3 and 6.  If the amount of
extinction is in fact of such magnitude, the consequence will be
profound: this much dust is capable of completely extinguishing the
light from the QSO in the restframe UV, allowing us to see the
underlying stellar population in high redshift QSOs for the first
time.

Given a high extinction to the QSO, the origin of the UV continuum
light is unclear.  It becomes especially puzzling if we consider the
fact that the spectra of these objects are as flat as those of
normal QSOs ($f_{\nu} \propto \nu^{-1}$, see Fig.~\ref{fig3}).  In
other words, {\em these objects have simply too much UV light for
their Balmer decrement values.}  In what follows, we will try
to explain the origin of the observed UV light in terms of three
different hypotheses: reddened QSO light, scattered QSO light, and 
star light from the surrounding stellar population.

The main requirement for these models is to explain the two salient
characteristics of these QSOs: the large Balmer decrements and the
small equivalent widths of the BELs, especially those of \mgii\
(\cite{Cowie94b}) and \hb\ (Table~1).  We do not include the
high-ionization BELs here because their small equivalent widths could
be explained by other mechanisms.  For example, large blueshifts of
high-ionization BELs, often seen in QSOs, will result in the
destruction of these lines by the broad absorption line region (BALR).
In the case of Q0059-2735, a blue shift of $\sim 700 $ km s$^{-1}$ was
suggested for the \civ\ BEL by Wampler, Chugai, \& Petitjean (1995).
On the other hand, if most high-ionization BELs actually come from the
BALR as suggested by Turnshek (1984), dust in the BALR, provided that
it is mixed with the ions, could preferentially destroy resonance-line
photons, which include most of the high-ionization BELs
(\cite{Voit93}).  Note, however, that neither of these explanations
would work with the \mgii\ BEL: a low-ionization BEL such as \mgii\ is
very unlikely to have a large blueshift, and also the \mgii\ BEL
probably cannot form in the BALR (\cite{Turnshek84a}).

In the following, we will examine in turn the three hypotheses
proposed above.  In order to estimate the contribution of the QSO
light, we assume its intrinsic spectrum to be: (1) $f_{\nu}
\propto \nu^{-1}$, (2) $\ha / \hb = 5$.

\subsection{The Reddened-QSO-light Hypothesis}
The reddened-QSO-light hypothesis explains the UV spectrum simply as a
reddened QSO continuum.  The upper plot of Fig.~\ref{fig3} shows
that if the flat continuum spectrum from a QSO (A) is reddened with
the SMC extinction law (\cite{Prevot84}) and $E(B-V)=0.08$, it can
more or less reproduce the overall shape of the observed spectrum of
Hawaii 167 (B).  This combination of the SMC extinction law with a
moderate amount of reddening was used to explain the SEDs of
low-ionization BALQSOs by Sprayberry \& Foltz (1992).
This might also be the case with Q0059-2735, but it is hard to judge
only from the broad-band UV measurements we have (lower plot).

The problem with this explanation is its inevitable conclusion that
the continuum suffers much less reddening than the broad Balmer lines
($E(B-V)=0.08$ vs.\ 0.70 for Hawaii 167).  It is almost certain that
the dust responsible for the reddening does not exist in the broad
emission line region (BELR): the space between the BEL clouds is
exposed to continuum radiation too strong for the existence of dust
while the BEL clouds themselves seem to be free from internal dust,
showing no reduction of resonance BELs (e.g., \civ ) with respect to
non-resonance BELs (e.g., \ciii ) (\cite{Netzer90}).  The theoretical
calculation by Laor \& Draine (1993) also indicates that the dust
cannot probably exist in the BELR.  This means that the reddening must
take place outside the BELR, and probably within or behind the BALR
where dust is protected from the strong radiation.  However, it is
very difficult to imagine that such a dust distribution could affect
the BEL photons and the continuum photons differently, causing a
different amount of reddening for each component.  Therefore in this
case, we need to assume that the large Balmer decrements are intrinsic
to these QSOs, and that the actual reddening with the BELs is as small
as that of the continuum.  The problem is that it is unclear whether
such a large deviation of the intrinsic Balmer decrements is feasible.
Also, this still cannot explain the small equivalent widths of \mgii\
and \hb\ because reddening by itself cannot change the equivalent widths
of the BELs.

\subsection{The Scattered-QSO-light Hypothesis}
The scattered-QSO-light hypothesis would explain the UV light as
scattered light from the QSO while maintaining the same amount of
reddening for both the BELs and the continuum.  In the
Fig.~\ref{fig4}, the flat QSO spectra (A and C) are reddened by the
SMC extinction law with color excesses derived from the Balmer
decrements, assuming its intrinsic value of 5.  These reddened QSO
spectra are then scaled such that they would give the observed flux at
the \ha\ wavelength (B and D).  Note that such a scaling will minimize
the amount of the UV light which must be explained as scattered light.
With this procedure, the contribution of the reddened QSO light to the
observed flux at \mgii , \hb , and \ha\ is estimated to be 13\%, 58\%,
and 100\% for Hawaii 167, and 24\%, 88\%, and 100\% for Q0059-2735,
respectively.  In other words, most of the UV light must be scattered
light.  The contribution from the scattered light component must then
be such that if combined with the reddened continuum, it will produce
an apparent spectrum which matches a flat spectrum with moderate
reddening.

The problem with this model is the extreme efficiency of scattering
required to produce the observed UV spectrum.  This is immediately
clear from the fact that the observed spectra of these objects are as
flat as those of normal QSOs.  In the case of Hawaii 167, the observed
flux at 2000 \AA\ is still 35 \% of the original power-law value.
Although such an efficiency is possible for a single-scattering
process, it is still impossibly high for a process involving a large
number of scatterings, which is undoubtedly the case for this heavily
reddened QSO.  If we assume that each photon moves outward by random
walks, it must go through on the order of $(\tau_{2000})^{2}$
scatterings, where $\tau_{2000}$ is the optical depth at 2000 \AA .
Since A(2000 \AA ) $=$ 6.6 for $E(B-V)=0.70$ and $R=3.1$ with the SMC
extinction law, $\tau_{2000}$ is $\sim 6.1$. Therefore, each photon
must be scattered $\sim$ 37 ($=6.1^{2}$) times before escaping
outside.  In order to achieve an effective efficiency of 35\% after
this many scatterings, we need a dust albedo of 0.97 at 2000 \AA ,
which is quite unlikely.  Although it is certain that there exists
some amount of scattered QSO light, there is probably not so much as to
explain all the UV light seen.  This model also cannot explain the
small equivalent widths of \mgii\ and \hb\ because scattering by
itself cannot change the equivalent widths of the BELs.

So far, we have seen that neither the reddened-QSO-light hypothesis
nor the scattered-QSO-light hypothesis can explain the nature of these
objects very well.  The main problem is the mismatch between the large
Balmer decrements and the flat spectra of these objects.  Furthermore,
neither hypothesis can explain the observed small equivalent widths of
the \mgii\ and \hb\ BELs.  For these models to be true, we need to
assume that either their small equivalent widths are intrinsic to the
QSOs, or that there is some mechanism at work which preferentially destroys
the BEL photons.  However, a more natural explanation would be to
invoke another component of the light which is diluting the QSO light.
This leads to the possibility that the UV spectra of these QSOs are
dominated by the light from their surrounding stellar populations.  We
now examine this hypothesis.

\subsection{The Star-light Hypothesis}
Direct support for the star-light hypothesis comes from the suggested
detection of the 4000 \AA\ break in the spectrum of Hawaii 167.
Although the shape of the break is ambiguous, there is clearly an
offset in the continuum level between the $J$ and $H$ bands of Hawaii
167 (Fig.~\ref{fig1}).  In Fig.~\ref{fig4}, we tried to fit this
4000 \AA\ break with the isochrone-synthesis model by Bruzual and Charlot
(1993) after subtracting the reddened-QSO-light model C from the
observed spectrum.  One thing is immediately obvious: the small 4000
\AA\ break is incompatible with the large drop of flux around 2000
\AA.  A young stellar population consistent with the small 4000 \AA\
break produces too much UV light (A) while an older population
matching the flux drop at 2000 \AA\ becomes too red in the UV--optical
color (B).  Here, we used the instantaneous starburst model to
accelerate the depression of the UV light as much as possible, but it
was still not fast enough to produce these two features
simultaneously.  One way to get around this problem is to introduce a
moderate amount of reddening.  An extremely young flat-spectrum
population (D) reddened with the SMC extinction law and $E(B-V)=0.23$
can more or less reproduce the observed spectrum (E).  Although the
fit is not perfect, it is probably acceptable if we consider the
various uncertainties such as the geometry of the dust distribution
and the properties of the dust particles, and in any case as good as
the line (B) in Fig.~\ref{fig3}.  Fig.~\ref{fig5} shows the
comparison between the model (E) and the actual near-IR spectra shown
in Fig.~\ref{fig1}.

A caveat to this explanation is its implicit but unavoidable
conclusion that the BALs seen in these QSOs have nothing to do with
the BALQSO phenomena, which we usually associate with the activities of
the QSO itself, but are features in the starbursting galaxy
surrounding the QSO.  Recall our argument that the reddening of the
QSO continuum and the BELs should take place within or behind the BALR
where dust is protected from the strong radiation (see Section 3.1).
If this is the case, whatever is lying inside the BALR must be as
reddened as the continuum and the BELs.  Since the stellar UV
continuum seems to suffer much milder reddening, this will put the
position of the stellar population outside the BALR.  However, such a
configuration obviously cannot produce the BALs in the stellar UV
continuum.  Therefore, we need to think that the BALs seen in these
QSOs are produced by this same stellar population.  In the case of
Hawaii 167, the widths of the BALs are somewhat narrower than usual
($\sim$ 1500 (\mgii ) -- 4500 (\siiv) km s$^{-1}$), and therefore they
might be explained as stellar absorption lines produced by a
starbursting population (\cite{Cowie94b}); The BALs seen in
Q0059-2735, on the other hand, show outflow velocities as high as
20000 km s$^{-1}$, so the ejecta of supernova explosions are the only
possible explanation.

This model, though plausible, cannot help but look somewhat arbitrary
in that it devises yet another mechanism for producing the BALs.
Here, we suggest that it is possible to construct a more simple and
coherent model of these QSOs and the BALQSO phenomena in general in
the framework of the dusty nuclear-starburst model (\cite{Lipari94a};
\cite{Lipari94}).  Suppose that in the nuclear region of these
galaxies, the central BELR is surrounded by a shell of starbursting
stellar population mixed with dust.  The BELs will show large amounts
of reddening because the BEL photons must transverse this dusty shell.
The stellar continuum, on the other hand, will show much smaller
amounts of reddening partly because the composite light is dominated
by stars with smaller extinction, and partly because scattered light
compensates for reddening by extinction (\cite{Witt92}).  The BALs are
probably produced in this shell as a result of supernova explosions.
The ejecta of these explosions will preferentially extend outward and
show blueshifts because of the smaller ambient density toward that
direction.  We see these outflowing gas clouds against the stellar
continuum or the QSO continuum in the case of normal BALQSOs,
recognizing them as the BALs.

This model could also explain the main characteristics of the
low-ionization BALQSOs (\cite{Lipari94a}; \cite{Lipari94}).  For
example, such dusty environments will prevent ionizing photons from
escaping to the outer low-density regions where the formation of
\oiii\ is possible, explaining the absence or the extreme weakness of
this line.  Also, supernova explosions are a natural way to explain
the formation of Fe, whose emission and absorption lines are found to
be abnormally strong (\cite{Hazard87}; \cite{Wampler95}; see also 
Fig.~\ref{fig2}).  We further speculate that this shielding of
ionizing radiation also allows the existence of low-ionization ions
such as \mgii\ and \aliii , whose absorption lines characterize the
spectra of these QSOs.

These dust-enshrouded QSOs will eventually emerge from the dust
cocoons once the dust in the surrounding shell is cleared away by
various energetic processes such as radiation pressure from the QSOs,
supernova explosions, and stellar winds (\cite{Sanders88};
\cite{Lipari94a}; \cite{Lipari94}).  If this happens, all the
characteristics of these QSOs mentioned above will go away: the \oiii\
emission will become stronger because the ionizing photons start to
reach the outer low-density region; the low-ionization BALs will
disappear because these ions will be more highly ionized; and the
emission and absorption by \feii\ and \feiii\ will weaken because
these ions will also move to higher ionization stages.  The resultant
objects will be normal-looking BALQSOs.  Therefore, if this model is
correct, we expect that the low-ionization BALQSOs and the normal
BALQSOs form an evolutionary sequence.  Although the argument so far
does not require the geometry of the standard AGN Unified Theories, in
this framework such a transition might be identified as the
development of ionization cones while we are looking at these objects
pole-on.

\section{CONCLUSIONS}
Although any conclusion drawn from such a small data set is
necessarily tentative, at this point we think that the star-light
hypothesis is the most likely explanation for the spectra of these
QSOs.  It is very difficult to imagine that the intrinsic Balmer
decrement could be close to, or even larger than 10 as in the case for
the reddened-QSO-light hypothesis; nor is it easy to imagine that any
dust grains could have UV albedos close to unity as in the case for
the scattered-QSO-light hypothesis.  Furthermore, neither of these
could explain the small observed equivalent widths of the \mgii\ and
\hb\ BELs.  Therefore, we conclude that these objects are in fact heavily
dust-enshrouded young QSOs whose internal reddening is so large as to
completely extinguish the QSO light in the restframe UV, allowing us
to see the underlying stellar population.  

A possible counter argument to this explanation, nevertheless, is to
attribute the large Balmer decrements and the small equivalent widths
to the scatter in the intrinsic QSO properties.  This argument is
especially strong for Q0059-2735, whose Balmer decrement and \hb\
equivalent width are deviant but might still be within the range of
their natural distribution, which itself has a large scatter.  Our
response to such an argument is that low-ionization BALQSOs are very
likely to be dusty from other lines of evidence (see Introduction),
and therefore it is very natural to explain the peculiar properties of
these objects as the effect of dust.  Probably most of the
low-ionization BALQSOs discovered so far are not as heavily reddened
as the two objects studied here because the determination of the low
internal reddening by Sprayberry \& Foltz (1992) stems not only from
the continuum shape but also from the emission line strength.  Our
claim, however, is that there are objects which are extreme versions
of these optically selected low-ionization BALQSOs, and they are
probably as dusty as the ultraluminous IRAS galaxies.  The
identification of one of the ultraluminous IRAS galaxies,
PC07598+6508, as a low-ionization BALQSO (\cite{Lipari94}) strongly
supports this idea.  In this sense, these two classes of objects may
well be the same things looked at from different angles or at
different times.

A model of a young QSO surrounded by a shell of starbursting
population mixed with dust seems to explain (at least qualitatively)
the main characteristics of these QSOs, and possibly suggest the
evolutionary connection between the low-ionization BALQSOs and normal
BALQSOs.  This starburst in the shell surrounding a QSO could be
identified as spheroid formation (\cite{Cowie94b}; \cite{Kormendy92})
though this is a speculative guess at this point.  This model of
course has its own problems, and we mention a few here: First, it is
unclear whether we could really explain the BALQSO phenomena with
supernova explosions in this surrounding shell.  Second, a great deal
of fine-tuning seems to be required if we are to produce an almost
perfectly power-law SED out of two completely separate light
components, a QSO and its underlying stellar population.  Finally, it
is uncertain whether the use of the Bruzual-Charlot model is justified
for these objects.  The starburst might be limited to a class of
massive stars within a small mass range; also, the light from
supernovae might be dominating the SEDs.

We also mention another rather technical problem: that is, the
assessment of effects due to \feii\ emission features.  It is known that
this type of QSO shows strong \feii\ emission, and that these features are
especially conspicuous around \hb.  Therefore, it is possible that
the small equivalent width of \hb\ is simply due to a large
contribution of \feii\ emission to the continuum.  Also, our
measurement of the 4000 \AA\ break amplitude might be affected by the
\feii\ emission features.  Although these effects are potentially
serious, we are not able to assess the effects of \feii\ emission here
because our limited spectral coverage and resolution prevent us from
performing detailed modelling of \feii\ emission features.

Despite the various uncertainties, these objects present two very
interesting possibilities: that is, 1) heavy internal reddening of
QSOs could result in a separation of their stellar light and AGN light
in spectral space, and 2) if such heavily reddened QSOs exist at
high redshifts, they could easily have escaped our detection
until now.

The first point leads us to the possibility that we may be able to
study the underlying stellar populations of QSOs by using this type of
object.  In this paper, we have presented an example of such analyses,
which suggests that the stellar population in Hawaii 167 looks very
young, probably around 15 Myrs old if we take the fit of the
Bruzual-Charlot model at its face value.  The great importance of such
a spectral analysis is that we do not have to spatially resolve the
QSO: at high redshifts where such a spatial separation becomes
difficult, spectroscopic studies of this class of QSOs may be the only
way to understand what high-$z$ QSOs really are.

Regarding the second point, if this type of object exists in 
significant numbers, it is very likely that many of them would not have
been picked up in the previous optical surveys due to their faint UV
continuum and absence of strong UV emission lines.  In fact, Hawaii
167-like objects could be completely dark in the observer's optical
band if starbursts have not started in the host galaxy.  This implies
that there might be a whole population of dust-enshrouded young QSOs
at high redshifts which has so far eluded our detection.  We were able
to identify Hawaii 167 simply because it has barely enough UV
continuum for our optical spectroscopy with a 4-m class telescope, and
this might explain the flatness of Hawaii 167's SED as a selection
effect.  In any case, if these objects are in fact young QSOs emerging
from their dust cocoons, they are probably much more abundant at $z >
2$, where the comoving space density of QSOs is rapidly increasing.
The redshift of Hawaii 167 is indeed just above 2.  Wide-field
sensitive near-IR surveys should eventually tell us if such a
population of objects exists.

We thank S.\ Charlot for providing the isochrone synthesis model, and
E.\ M.\ Hu for helpful comments on the manuscript.  This work was
partly supported by the Grant-in-Aid of the Ministry of Education,
Japan (07044080).

\clearpage

\figcaption[fig1.ps]{Near-IR spectra of Hawaii 167.  The upper
spectrum was taken with the OH-airglow suppressor spectrograph while
the lower one was taken with the CGS4 spectrometer.  The thick lines
show the fits to the \ha\ and the \hb\ emission lines, using a
gaussian and a linear continuum.  The horizontal dash-dot lines
indicate the broad-band magnitudes listed in Cowie et al. (1994b).
Note the suggested detection of a 4000 \AA\ break in the upper plot.
The black solid circles indicate the flux of the fitted linear
continuum at the central wavelengths of the $J$, $H$, and $K^{\prime}$
bands.  These points are used in Figs.~3--5 as reference points when
the observed spectra are compared with the models.  The point at 4000
\AA, which lies on the extension of the $H$-band continuum fit, is
also used in the figures to see whether the models fit the possible 4000
\AA\ break feature.  Also, we conclude that the feature at $\sim$ 5007
\AA\ is an instrumental artifact because its width is narrower than the
instrumental profile. \label{fig1}}

\figcaption[fig2.ps]{Near-IR spectra of Q0059-2735.  Both spectra
were taken with the CGS4 spectrometer.  The thick lines are the fits
to \ha\ and \hb .  Two gaussians were used simultaneously for \ha\ to
improve the fit.  For H$\beta$, we mirror-reflected the left half of
the line to the right and then fitted a gaussian, in order to avoid
contamination from the \feii\ emission lines clearly seen to the
right. The horizontal dash-dot lines indicate the broad-band
magnitudes listed in Cowie et al. (1994b).  The black solid circles
indicate the flux of the fitted linear continuum at the central
wavelengths of the $J$ and $H$ bands.  These points are used in Figs.\
3 and 4 as reference points when the observed spectra are compared
with the models.
\label{fig2}}

\figcaption[fig3.ps]{
(A) A flat spectrum ($f_{\nu} \propto \nu^{-1}$) simulating the QSO
continuum for Hawaii 167; (B) The spectrum (A) reddened with the SMC extinction law
and $E(B-V) = 0.08$; (C) A flat spectrum QSO continuum simulation as in (A),
but for Q0059-2735.  The squares and
the triangles indicate the observed spectra for the two objects.  These spectra were
corrected for the foreground Galactic reddening by assuming
$E(B-V)=0.14$ for Hawaii 167 and $E(B-V)=0.02$ for Q0059-2735.  The
vertical error bars indicate a conservative estimate of 30\%
photometric errors.  
The broad-band measurements from
Cowie et al. (1994b) are shown as points with horizontal bars, whose lengths
correspond to the band widths, while the points without horizontal bars
are the reference points derived from our spectral measurements (black solid
circles marked in Figs.\ 1 and 2).  The optical spectrum
of Hawaii 167, taken from Cowie et al. (1994b), is scaled so that it
gives the correct $I$ magnitude as listed in that paper.  All
reference SEDs (A, B, and C) are scaled such that they match the observed
flux at the longest wavelength.
\label{fig3}}

\figcaption[fig4.ps]{
(A) A flat spectrum ($f_{\nu} \propto \nu^{-1}$) simulating the QSO
continuum for Hawaii 167; (B) The spectrum (A) reddened with the SMC extinction law
and $E(B-V) = 0.7$; (C) A flat spectrum QSO continuum simulation as in (A),
but for Q0059-2735; (D) The spectrum
(C) reddened with the SMC extinction law and $E(B-V) = 0.35$.  An
intrinsic Balmer decrement of 5 was assumed.  The SEDs are scaled
such that the reddened continuum matches the observed flux at \ha .
All symbols are the same as for Fig.~3.  The percentages
below the two solid lines indicate the relative contributions of (B) and (D) to
the observed spectra at the wavelengths of \mgii , \hb , and \ha .  In
the case of \mgii\ in Q0059-2735, this value is with respect to the
power-law line (C) in Fig.~3.  The percentage (35\%) just below the
UV spectrum of Hawaii 167 indicates the amount of the observed UV light
at 2000 \AA\ with respect to the power-law continuum (A).  Since the
amount of the reddened light (B) at this wavelength is negligible,
this is equal to the amount of scattered light.
\label{fig4}}

\figcaption[fig5.ps]{
(A) The Bruzual-Charlot instantaneous burst model with a Salpeter
IMF ($0.1<{\rm M}<125 {\rm M}_{\odot}$), and age = 38 Myr; (B) Same as for (A)
but with an age = 1 Gyr; (C) The estimated QSO continuum as for (B) in Fig.\ 
4 but scaled by assuming the intrinsic \hb\ equivalent width of 100
\AA . (D) The model SED shown in (A) but with an age = 15 Myr; (E) the spectrum (D) reddened with
the SMC extinction law and E(B--V)=0.23.  The squares in the upper
plot indicate the observed spectrum shown in Figs.~3 and 4 while the
solid circles show the spectrum after the subtraction of the QSO
spectrum (C).  The percentages indicate the contribution of (C) to the
observed spectra at the wavelengths of \mgii , \hb , and \ha .
\label{fig5}}

\figcaption[fig6.ps]{Model E from Fig.~4 is compared with the
observed near-IR spectrum of Hawaii 167 in Fig.~1 after correction of the observed
spectrum for the Galactic reddening. \label{fig6}}
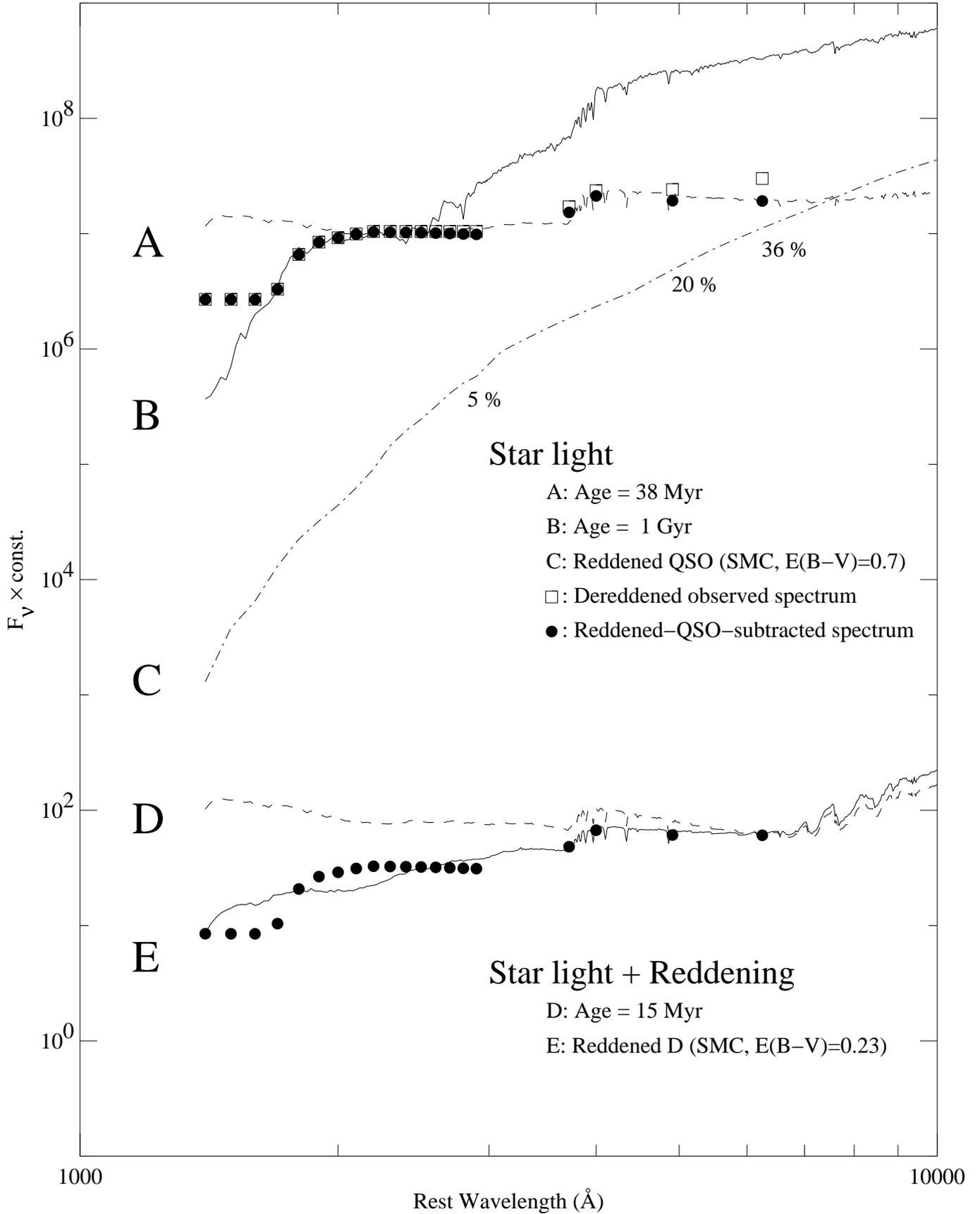
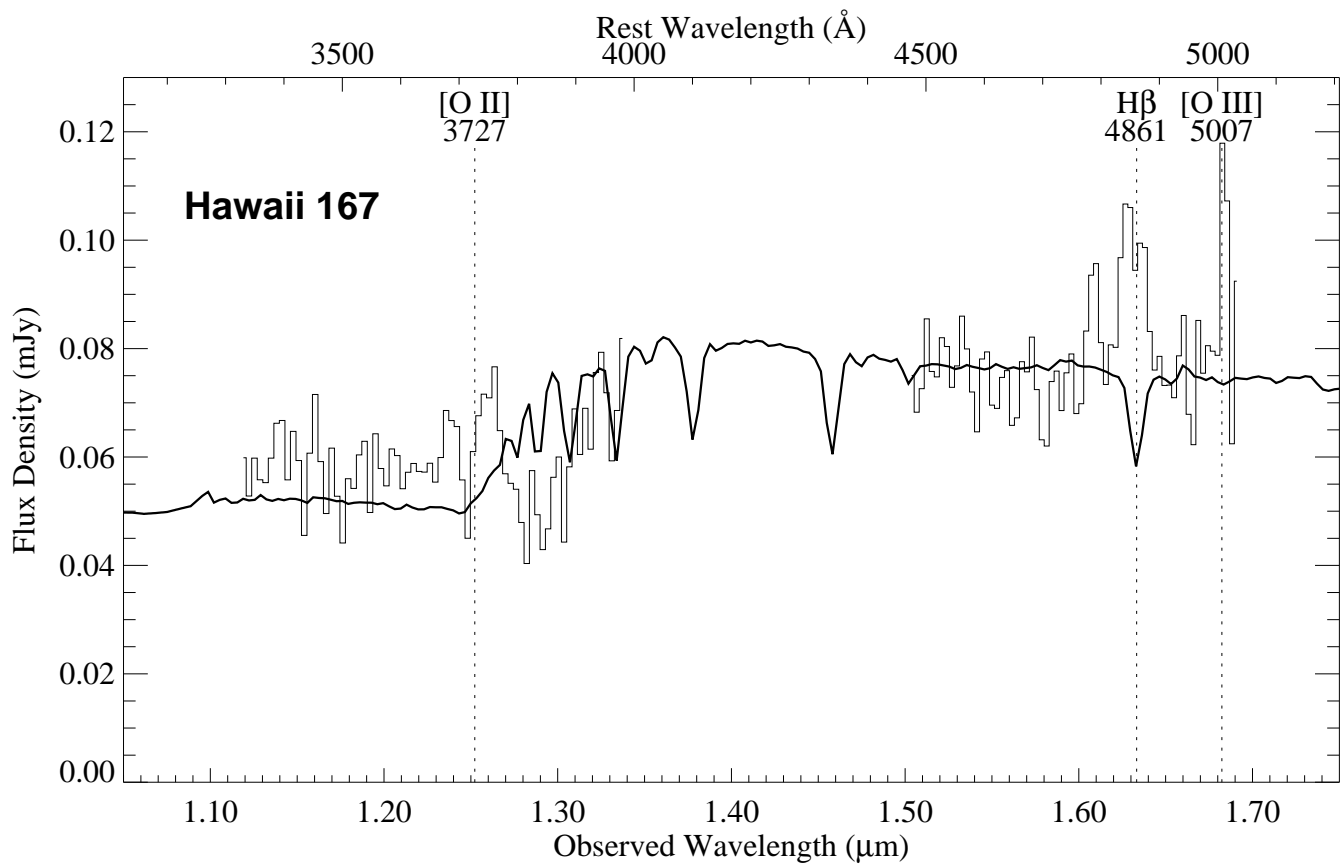
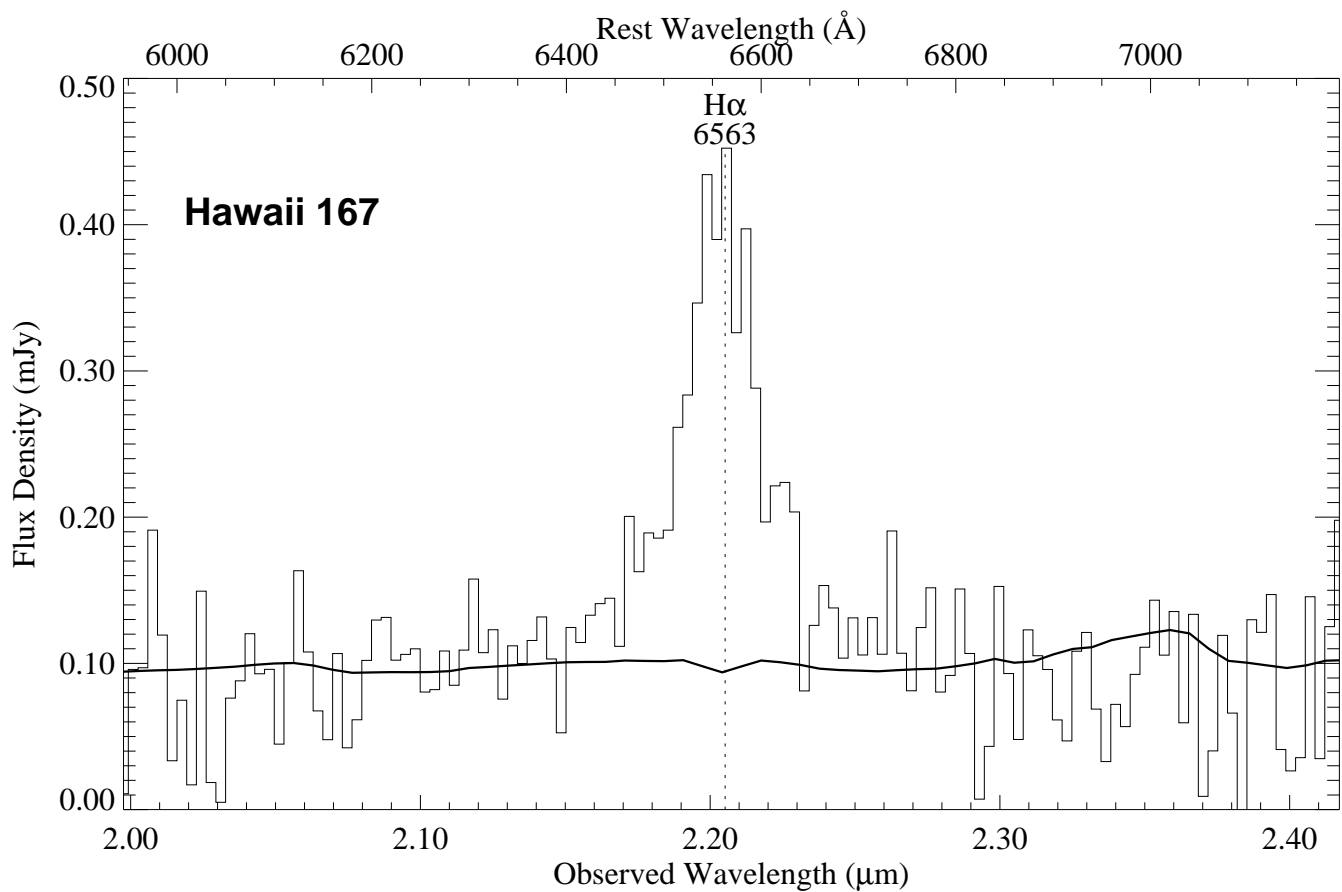
\end{document}